\documentclass{article}
\usepackage{spconf,amsmath,graphicx,hyperref}
\usepackage{svg}
\usepackage{booktabs, xcolor, makecell}
\usepackage{booktabs}
\usepackage{graphicx}
\usepackage{cite}
\usepackage[table]{xcolor}
\usepackage{float} 
\usepackage{adjustbox} 
\usepackage{indentfirst}
\usepackage{amsmath}
\usepackage{amssymb}
\usepackage{tabularx,array,xcolor,booktabs}
\newcolumntype{Y}{>{\centering\arraybackslash}X}
\usepackage{multirow}   
\usepackage{booktabs}   
\usepackage{colortbl}   
\usepackage{tabularx} 

\usepackage{xcolor}

\title{TICL+: A case study on Speech In-Context Learning for Children's speech Recognition}
\name{Haolong Zheng, Yekaterina Yegorova, Mark Hasegawa-Johnson}
\address{ University of Illinois at Urbana-Champaign
\\\{haolong2, yay2, jhasegaw\}@illinois.edu}
\begin{document}
%
\maketitle
\begin{abstract}
Children’s speech recognition remains challenging due to substantial acoustic and linguistic variability, limited labeled data, and significant differences from adult speech. Speech foundation models can address these challenges through Speech In-Context Learning (SICL), allowing adaptation to new domains without fine-tuning. However, the effectiveness of SICL depends on how in-context examples are selected. We extend an existing retrieval-based method, Text-Embedding KNN for SICL (TICL), introducing an \textit{acoustic reranking} step to create TICL+. This extension prioritizes examples that are both semantically and acoustically aligned with the test input. Experiments on four children’s speech corpora show that TICL+ achieves up to a 53.3\% relative word error rate reduction over zero-shot performance and 37.6\% over baseline TICL, highlighting the value of combining semantic and acoustic information for robust, scalable ASR in children’s speech.
  
\end{abstract}

\section{Introduction}
\label{sec:intro}
More than half of the children under the Individuals with Disabilities Education Act (IDEA) require speech and language services, which is approximately 3.4 million children.
Children with speech or language-related concerns risk falling behind in their academic and social-emotional development \cite{AI4ExceptionalEdTech}.
Typically, the earlier these concerns can be identified and addressed with ability-based interventions, the greater the likelihood that these children will thrive academically and socio-economically.
However, due to the substantial imbalance between the number of Speech and Language Pathologists (SLPs) and the children who require their services, there has been growing interest in automating these tasks to improve the efficiency of screening for language disorders. \cite{liu2024automatic}.

The success of such automation depends heavily on the accuracy and robustness of the automatic speech recognition (ASR) systems integrated into these pipelines.
ASR for children’s speech remains a low-resource task and exhibits a notable performance gap when applying off-the-shelf ASR systems directly due to the substantial acoustic and linguistic variability inherent in children’s speech, including inter-speaker variability due to differing developmental rates and intra-speaker variability resulting from underdeveloped pronunciation skills \cite{koenig2008speech,koenig2008stop,lee1999acoustics,lee1997analysis,vorperian2007vowel,smith1992relationships}. The resulting performance degradation is significant, as these sources of variability are largely absent from the data used to train large-scale ASR models. To address this challenge, transfer learning techniques have been employed to apply knowledge from adult ASR systems to children’s ASR, specifically by fine-tuning models such as Whisper \cite{attia2023kid,jain2023adaptation} and Wav2Vec2 \cite{jain2023wav2vec2} with children's speech data.
To mitigate data bias when fine-tuning self-supervised learning models with data from a different domain than the pretraining data, \cite{fan2022draft} proposed a Domain-Responsible Adaptation and Fine-Tuning strategy and reported improvements in word error rate (WER) across multiple speech models when fine-tuned with a children's speech dataset \cite{9864219,fan2022draft}.

\begin{figure}[!th]
    \centering

    \includegraphics[width=1.05\linewidth]{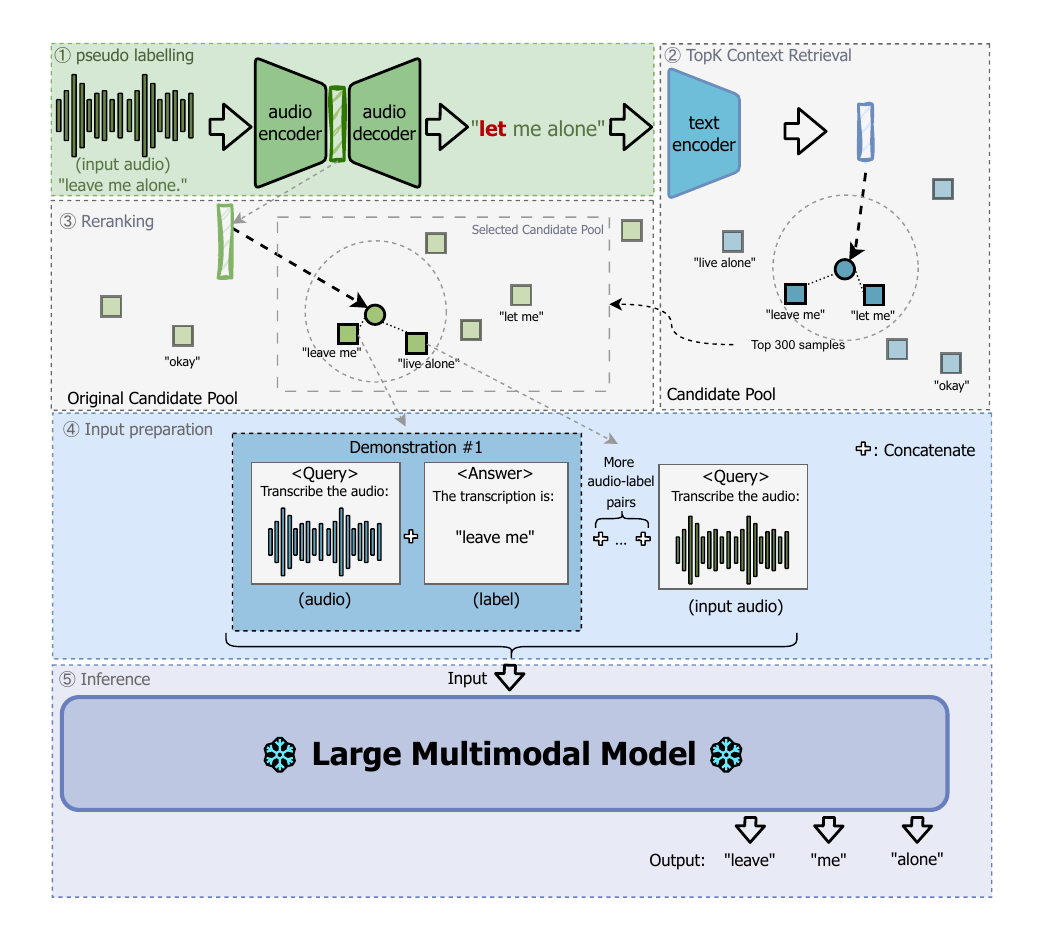}
    \label{fig:pipeline}
    \caption{Overview of the TICL+ pipeline}
\end{figure}

Beyond fine-tuning methods, in-context learning (ICL) \cite{NEURIPS2020_1457c0d6,dong-etal-2024-survey} has emerged as a flexible adaptation paradigm for large language models (LLMs) that mitigates catastrophic forgetting and eliminates the need for parameter updates.  SICL studies rely on random sampling to select in-context examples \cite{10446502, roll2025incontextlearningboostsspeech, zhou2025m2rwhispermultistagemultiscaleretrieval}, however, previous work has demonstrated that the selection of in-context examples strongly influences the performance of ICL \cite{pmlr-v139-zhao21c, yang-etal-2023-representative, NEURIPS2024_8cb564df}. To make the selection of in-context examples more targeted, we previously introduced a Text-
Embedding KNN for SICL (TICL) pipeline that first generates a pseudo-label for the test sample and then retrieves semantically similar demonstrations, enhancing SICL performance \cite{zheng2025ticl}. This method is dependent on the pseudo-labels, whose quality can be substantially degraded in low-resource scenarios such as children’s speech. In these settings, incorporating acoustic similarity as an additional similarity measure can prove to be beneficial. To address this, we extend the retrieval stage of TICL by introducing an acoustic-based reranking step that prioritizes demonstrations with acoustic characteristics closer to the test utterance, resulting in the proposed \textbf{TICL+} pipeline. This dual-criteria selection strategy, illustrated in Fig.~\ref{fig:pipeline}, improves context construction for SICL in low-resource speech domains such as children’s speech.


\section{Methodology}
\label{sec:methodology}

\subsection{Speech In-Context Learning}
\label{sec:SICL}
Rather than updating the model parameters, ICL adapts a model to a target domain by conditioning on demonstrations drawn from the target domain. SICL extends text-based ICL by conditioning jointly on paired audio and text tokens. Given a test speech sample $\boldsymbol{s^*}$, a model $\Lambda$ generates a transcription $\boldsymbol{\hat{y}}$ conditioned on context $C$:

\begin{equation*}
    \boldsymbol{\hat{y}} = \arg\max_{\boldsymbol{y}} \mathrm{Pr}(\boldsymbol{y} \mid C, \boldsymbol{x_s^*}, \Lambda),
\end{equation*}
where $\boldsymbol{x_s^*}$ denotes the audio encoding of $\boldsymbol{s^*}$.

The context $C$ consists of query–answer pairs $c^{(i)}=(q^{(i)}, a^{(i)})$, where each query $q^{(i)}$ is an encoded audio segment, and each answer $a^{(i)}$ corresponds to the transcription of that audio segment.

\subsection{Text-Embedding KNN Candidate Selection}
\label{sec:TICL}
The TICL pipeline \cite{zheng2025ticl} introduces a text-embedding-based KNN candidate selection method designed to identify an effective context $C$ for SICL for a given test sample. To construct $C$, TICL retrieves speech-transcription pairs whose transcriptions are lexically similar to the test utterance from a candidate dataset $\mathcal{C}=\{(\boldsymbol{s^{(i)}}, \boldsymbol{y^{(i)}})\}_{i=1}^N$ where $\boldsymbol{s^{(i)}}$ denotes the speech audio and $\boldsymbol{y^{(i)}}$ its corresponding transcription. 

A frozen text encoder $\phi: \mathcal{Y}_{\mathrm{text}} \rightarrow \mathbb{R}^d$ maps each transcription $\boldsymbol{y}$ to a $d$-dimensional sentence embedding. The $\ell_2$-normalized embedding is defined as:  

\begin{equation*}
\bar{\phi}(\boldsymbol{y})=\frac{\phi(\boldsymbol{y})} {\|\phi(\boldsymbol{y})\|_2}
\end{equation*}
For each candidate $c_i \in \mathcal{C}$, its normalized embedding is precomputed as
$\bar{\boldsymbol{z}}^{(i)} = \bar{\phi}(\boldsymbol{y^{(i)}})$. During inference, the ground-truth transcription of the test utterance $\boldsymbol{s^*}$ is unavailable. Instead, a pseudo-transcription $\boldsymbol{\tilde{y}} = f_\theta(\boldsymbol{s^\ast})$ is generated using a frozen ASR model $f_\theta: \mathcal{X}{_\mathrm{audio}} \rightarrow \mathcal{Y}{_\mathrm{text}}$, where $\mathcal{X}{_\mathrm{audio}} ,\mathcal{Y}{_\mathrm{text}}$ denote the audio and the text space respectively.

The pseudo-label is then encoded into a normalized lexical embedding $\boldsymbol{\bar{z}^\ast}=\bar{\phi}(\boldsymbol{\tilde{y}})$. To select relevant in-context examples, we compute the Euclidean distance between $\bar{\boldsymbol{z}}^*$ and each candidate embedding $\bar{\boldsymbol{z}}^{(i)}$: 

\begin{equation*}
r(i) \;=\; \bigl\|\boldsymbol{\bar{z}^\ast} - \boldsymbol{\bar{z}^{(i)}}\bigr\|_2.
\end{equation*}
The $K$ most similar candidates are retrieved as 

\begin{equation*}
\mathcal{N}_K(\boldsymbol{s^\ast})\;=\;\operatorname{TopK}_{i\in[N]}\!\bigl(-r(i)\bigr),
\end{equation*}
and used to construct the final context $C$.

TICL was evaluated on \texttt{Phi-4-MultiModal-Instruct} (Phi-4-MM)\cite{microsoft2025phi4minitechnicalreportcompact}.


\subsection{Acoustic Reranking for Refining Context Selection}

To better align the context $C$ with the acoustic characteristics of the test audio, we extend the TICL pipeline    to TICL+ by introducing an acoustic reranking step. Prior results demonstrated that Whisper embeddings were the second-best retrieval method after semantic similarity \cite{zheng2025ticl}. Whisper embeddings can capture many aspects of the input speech, such as prosody, speaker identity, and pronunciation, that are not reflected in purely lexical representations \cite{Gong_2023}. Motivated by this, we incorporate an acoustic-based distance measure to refine the selection of in-context examples.

Using the top $M$ semantically similar candidates $\mathcal{N}_M(\boldsymbol{s^*})$, where $M=300$, we compute acoustic similarity using precomputed embeddings using a frozen speech encoder $g: \mathcal{X}{_\mathrm{audio}} \rightarrow\mathbb{R}^p$. \texttt{Whisper-large-v3-turbo} was used in our experiments. The $\ell_2$-normalized acoustic embedding is precomputed as
\begin{equation*}
\label{eq:acoustic_precompute}
\bar{g}(\boldsymbol{s})= \frac{g(\boldsymbol{s})}{\|g(\boldsymbol{s})\|_2}.
\end{equation*}

The acoustic distance between the test audio and each candidate is then computed as
\begin{equation*}
\label{eq:acoustic_distance}
r_{\text{acoustic}}(i) = \|\bar{\boldsymbol{a}}^* - \bar{\boldsymbol{a}}^{(i)}\|_2.
\end{equation*}

Where $\bar{\boldsymbol{a}}^* = \bar{g}(\boldsymbol{s^*})$ and $\bar{\boldsymbol{a}}^{(i)} = \bar{g}(\boldsymbol{s^{(i)}})$. Candidates are reranked according to $r_{\text{acoustic}}(i)$, 
and the top $K$ acoustically closest samples are selected as:
\begin{equation*}
\label{eq:acoustic_reranking}
\mathcal{N}^{\text{acoustic}}_K(\boldsymbol{s^*})
= 
\operatorname{TopK}_{i \in \mathcal{N}_M(\boldsymbol{s^*})}
\!\bigl(-r_{\text{acoustic}}(i)\bigr).
\end{equation*}

The final SICL context $C$ is constructed from 
$\mathcal{N}^{\text{acoustic}}_K(\boldsymbol{s^*})$. 
This two-stage retrieval process ensures that the selected in-context examples are both lexically related and acoustically similar to the test utterance.


\section{Experimental Results and Analysis}
To evaluate children's speech recognition performance, we used four corpora: My Science Tutor (MyST) \cite{pradhan2024my}, a containing science tutoring dialogues with students ages 8–11; the OGI Kids' Speech Corpus \cite{ogi} which contains about 100 hours of read and prompted speech from children ages 5-16;  Edmonton Narrative Norms Instrument (ENNI) \cite{schneider2006storytelling, liu2024fasaflexibleautomaticspeech}, which contains narrative retellings and story completions by children ages 4-9; and the Redmond Sentence Recall (RSR) \cite{ai4exceptionaled_rsr_hf} consisting of sentence repetition tasks for children ages 5–9, including those with developmental language disorders. All datasets were preprocessed, and the corresponding candidate sets were generated according to the procedure outlined in \cite{zheng2025ticl}. We evaluate TICL+ using Phi-4-MM.

Table~\ref{tab:child-wer} demonstrates that incorporating acoustic reranking into the TICL pipeline substantially improves recognition accuracy across all datasets. TICL+ achieves up to a 53.3\% relative improvement over zero-shot performance and up to 37.62\% over TICL. These gains may stem from the limitations of the pseudo-labeler, which often produces inaccurate transcriptions for children’s speech. Selecting the top 300 semantically closest utterances helps remove unrelated examples, 
while the acoustic reranking step further refines the context by prioritizing samples that are acoustically similar to the test utterance, 
regardless of their lexical similarity.

Across all four datasets, TICL+ consistently outperforms both zero-shot and TICL. 
The largest relative improvement (53.3\%) is observed on MyST, which contains conversational speech with high variability in speaker age and background noise. This suggests that acoustic similarity may help identify examples with comparable speaker and environmental conditions, resulting in more robust contextual alignment. 

Smaller improvements are observed on OGI and ENNI, which primarily contain read or structured speech where lexical overlap already provides strong guidance during the semantic retrieval step. In these cases, the acoustic filter likely contributes by accounting for pronunciation differences and developmental variability across speakers. Performance on RSR also improves despite it also containing read utterances, showing that acoustic reranking remains effective even when lexical diversity is limited, potentially due to its ability to identify developmentally similar speakers.

Overall, these results demonstrate that acoustic similarity provides complementary information to lexical similarity, enabling the model to better capture inter- and intra-speaker variability inherent in children’s speech. The improvements across all corpora demonstrate that TICL+ is effective for low-resource and developmentally variable speech domains.

\begin{table}[!t]
\centering
\ninept
\caption{Children's Speech Results for \emph{Phi-4-MM}. \(\downarrow\)WER\%.}
\label{tab:child-wer}
\begin{tabularx}{\linewidth}{@{}l l YYYY@{}}
\toprule
\textbf{Method} & \textbf{$k$} & \textbf{MyST} & \textbf{OGI} & \textbf{ENNI} & \textbf{RSR} \\
\midrule
\textbf{Zero-Shot} 
& $0$ & 12.81 & 16.17 & 14.37 & 20.06 \\
\cmidrule[\heavyrulewidth]{1-6}
\multirow{5}{*}{\textbf{TICL}} 
& $1$ & 17.27 & 9.55 & 17.57 & 18.92 \\
& $2$ & 11.77 & 8.94 & 14.07 & 18.92 \\
& $3$ & 11.69 & 8.75 & 13.54 & 18.90 \\
& $4$ & 11.81 & 8.52 & 13.75 & 19.54 \\
\cmidrule[\heavyrulewidth]{2-6}
& \emph{$\Delta_{\mathrm{rel}}$} & 8.7\% & 47.3\% & 5.8\% & 5.8\% \\
\cmidrule[\heavyrulewidth]{1-6}
\multirow{6}{*}{\textbf{TICL+}} 
& $1$ & 11.48 & 8.84 & 14.83 & 12.89 \\
& $2$ & \textbf{10.17} & 7.97 & 12.01 & 12.26 \\
& $3$ & 10.52 & 7.78 & \textbf{11.52} & \textbf{12.19} \\
& $4$ & 10.57 & \textbf{7.55} & \textbf{11.52} & 12.75 \\
\cmidrule[\heavyrulewidth]{2-6}
& \emph{$\Delta_{\mathrm{rel}}$} & \textbf{20.6\%} & \textbf{53.3\%} & \textbf{19.8\%} & \textbf{39.2\%} \\
\bottomrule
\end{tabularx}
\end{table}

\section{Conclusion}
\label{sec:conclusion}

In this work, we introduced an acoustic reranking step into the TICL pipeline, leading to the TICL+ framework, which improves SICL for children's speech recognition. The proposed approach leverages both acoustic and semantic similarity to construct more effective in-context examples. Experiments on four children's speech corpora demonstrate significant performance gains, reducing relative WER by up to a 53.3\%  over zero-shot performance and up to 37.62\% over the baseline TICL. These findings highlight the value of incorporating multiple factors when selecting in-context examples, paving the way toward more robust and scalable ASR systems for children’s speech.

\section{Acknowledgments}
This work was supported by National Science Foundation grant \#2229873. 
This work used the Delta system at the National Center for Supercomputing Applications through allocation beiq-delta-gpu from the Advanced Cyberinfrastructure Coordination Ecosystem: Services \& Support (ACCESS) program, which is supported by National Science Foundation grants \#2138259, \#2138286, \#2138307, \#2137603, and \#2138296.

{\ninept
\bibliographystyle{IEEEbib}
\bibliography{strings}
}

\end{document}